\documentstyle[12pt,aasms4,epsf]{article}
\begin{document}

\title{ An Additional Application of the Space Interferometry\\ 
	Mission to Gravitational Microlensing Experiments}
\bigskip
\bigskip

\author{Cheongho Han}
\bigskip
\affil{Department of Astronomy \& Space Science, \\
       Chungbuk National University, Chongju, Korea 361-763 \\
       cheongho@astronomy.chungbuk.ac.kr}
\bigskip
\authoremail{cheongho@astronomy.chungbuk.ac.kr}
\bigskip
\bigskip
\author{Tu-Whan Kim}
\bigskip
\affil{Department of Astronomy, \\
       Yonsei University, Seoul, Korea 120-749}
\bigskip

\begin{abstract}
   Despite the detection of a large number of gravitational microlensing 
events, the nature of Galactic dark matter remains very uncertain.  This
uncertainty is due to two major reasons: the lens parameter degeneracy in the
measured Einstein timescale and the blending problem in dense field
photometry.  Recently, consideration has been given to routine astrometric
followup observations of lensing events using the {\it Space Interferometry
Mission} (SIM) as a means of breaking the lens parameter degeneracy in 
microlensing events.  In this paper, we show that in addition to breaking the
lens parameter degeneracy, SIM observations can also be used to correct for 
nearly all types of blending.  Therefore, by resolving both the problems of the 
lens parameter degeneracy and blending, SIM observations of gravitational 
lensing events will significantly better constrain the nature of Galactic dark 
matter.
\end{abstract}

\vskip25mm

\centerline{resubmitted to {\it Monthly Notices of the Royal Astronomy 
Society}: Dec.\ 28, 1998}
\centerline{Preprint: CNU-A\&SS-09/98}
\clearpage

\section{Introduction}

   Surveys to detect Galactic dark matter by monitoring light 
variations of source stars caused by gravitational microlensing are being
carried out by several groups and $\sim 300$ events have been detected
(MACHO: Alcock et al.\ 1997a, 1997b; EROS: Ansari et al.\ 1996; 
OGLE: Udalski et al.\ 1997).  The light curve of a lensing event is 
represented by
$$
A_{\rm abs} = { u^2+2\over u(u^2+4)^{1/2}};\qquad 
u = \left[ \beta^2 + \left( {t-t_0\over t_{\rm E}} \right)^2\right]^{1/2},
\eqno(1.1)
$$
where the lensing parameters $\beta$, $t_0$, and $t_{\rm E}$ represent the
lens-source impact parameter, the time of maximum amplification, and the
Einstein ring radius crossing time (Einstein timescale), respectively.  Once
the light curve of an event is observed, these lensing parameters are
obtained by fitting the theoretical light curves of equation (1.1) to the
observations.  One can obtain information about individual lenses because the
Einstein timescale is related to the physical lens parameters by
$$
t_{\rm E} = {r_{\rm E}\over v},\qquad 
r_{\rm E} = \left( {4GM\over c^2}{ D_{ol}D_{ls} \over D_{os}}\right)^{1/2},
\eqno(1.2)
$$
where $r_{\rm E}$ is the Einstein ring radius, $v$ is the lens-source
transverse speed, $M$ is the lens mass, and $D_{ol}$, $D_{ls}$, and $D_{os}$
are the separations between the observer, lens, and source star.  However,
since the Einstein timescale depends on a combination of the lens parameters,
the values of the lens parameters determined from it suffer from large
uncertainties.

   The lens parameters suffer from additional uncertainties due to blending.  
The probability for a single source star to be in the state of gravitational
amplification, i.e., the optical depth $\tau$, is very low; from $({\cal O})
10^{-7}$ to $({\cal O}) 10^{-6}$ depending on the observed field.  To
increase the event rate, therefore, experiments are being conducted toward
very dense star fields such as the Galactic bulge and Magellanic Clouds.
However, photometry of such dense star fields is affected by the flux from
stars not participating in the gravitational lensing, the ``blending problem.''
Depending on the origin of the blended light, blending is classified into 
regular, amplification-bias, binary-source, and bright-lens blending 
(see \S\ 2).  When an event is affected by blending, the observed light 
curve is represented by
$$
A_{\rm obs} = A_{\rm abs}(1-f) + f;\qquad f = { B\over F_0+B},
\eqno(1.3)
$$
where $F_0$ is the unblended flux of the source star before (or after)
gravitational amplification, $B$ is the amount of blended flux, and thus $f$
represents the fraction of the blended light to the total observed flux.
Therefore, when fitting a blended lensing event light curve, an additional
lensing parameter $f$ must be included.  As a result, the determined lens
parameters become even more uncertain.

   Efforts to improve the results of gravitational microlensing 
experiments are therefore focused on breaking the lens parameter degeneracy
and correcting for the blending effect.  Various methods have been proposed
to do this. The methods for removing the degeneracy in the lens parameters 
are summarized by Jeong, Han, \& Park (1999).  Here we discuss the methods 
proposed to correct for various types of blending (see \S\ 2).  However, these 
methods are either applicable to only in a few rare cases or impractical due to 
the large uncertainty in the determined value of $f$.  Therefore, to gain 
improved results from the next generation of lensing experiments, it is 
essential to devise practical methods that can resolve the problems of the 
lens parameter degeneracy and blending in general microlensing events.

   Recently, routine astrometric followup observations of lensing events 
with high precision instruments such as the {\it Space Interferometry
Mission} (hereafter SIM, Allen, Shao, \& Peterson 1998) are being discussed
as a method to break the lens parameter degeneracy of general microlensing
events.  When a source star is gravitationally lensed, it is split into two
images.  The image separation is too small for direct observation.  However,
the displacements of the light centroid of the two images can be measured
with SIM.  The trajectory of the centroid shifts is an ellipse (astrometric
ellipse) whose shape depends on the lens-source impact parameter 
(see \S\ 4.2).  The usefulness of astrometric measurements of lensing events 
is that one can determine the angular Einstein ring radius, 
$\theta_{\rm E}=r_{\rm E}/D_{ol}$, from the measured centroid shifts because 
the size of the astrometric ellipse is directly proportional to 
$\theta_{\rm E}$.  While the Einstein timescale depends on three lens 
parameters ($M$, $D_{os}$, and $v$), the angular Einstein ring radius depends 
only on two parameters ($M$ and $D_{ol}$).  Therefore, by measuring 
$\theta_{\rm E}$, the uncertainties in the determined lens parameters can 
be significantly reduced (Walker 1995; Paczy\'nski 1998; 
Boden, Shao, \& Van Buren\ 1998; Han \& Chang 1998).

   In this paper, we show that in addition to breaking the lens parameter 
degeneracy, observations with SIM can be used to correct for the blending
effect in general microlensing events.  Due to the high resolution, 
$\lesssim 0.\hskip-2pt ''5$, from the diffraction-limited space observations,
a significant fraction of events blended by field stars, i.e., regular and 
amplification-bias blended events, are automatically resolved by a single 
mirror of SIM.  For events affected by very close blends with 
separations $\Delta\theta \lesssim 10\ {\rm mas}$, including a large fraction of
binary companions and all bright lenses, one can also identify the lensed 
source star by detecting the distortions of the astrometric centroid 
trajectory caused by the blended star.  For events affected by medium range 
separation blends, 
$10\ {\rm mas}\lesssim\Delta\theta\lesssim 0.\hskip-2pt ''5$, the blending 
effect can be corrected for by either detecting the shift of the broad 
envelope centroid between the point spread functions (PSFs) of lensed and 
blended stars caused by gravitational amplification or by examining the 
narrow fringe patterns observed at multi-wavelengths.  Therefore, by 
resolving both the problems of lens parameter degeneracy and blending, SIM 
observations of gravitational lensing events will significantly better 
constrain the nature of Galactic dark matter.

\section{Types of Blending}
   Depending on the origin of the blended light, blending is classified into
several types.  First, ``regular blending'' occurs when a star brighter than
the detection limit, which is set by crowding, is lensed and the measured
flux is affected by the residual flux from other blended stars fainter than
the detection limit.  Due to regular blending, the apparent amplification of
the event is lower than its intrinsic value.  As a result, the apparent
Einstein timescale is shorter than its true value, resulting in systematic
underestimation of lens masses (Di Stefano \& Esin 1995; Wo\'zniak \&
Paczy\'nski 1997; Han 1998b).  Since the optical depth is directly 
proportional to the timescale, the value of $\tau$ determined without properly 
correcting for regular blending is also underestimated.

   The observed light curve is also affected by blending when one of several
unresolved faint stars below the detection limit is lensed and its flux is
associated with the flux from other field stars in the effective seeing disk 
of a monitored bright source star.  This is known as ``amplification-bias
blending'' (Bouquet 1993; Nemiroff 1994).  The effects of amplification-bias
blending for the determination of individual lens parameters are similar to
those of regular blending, but the effects are more severe due to the much
larger amount of blended light.  However, the most important effect of
amplification bias blending is that when not accounted for, the optical depth
might be significantly overestimated (Alard 1997).  This is because
$\tau$ is determined based only on the number of monitored stars brighter
than the detection limit, while events are actually detected among a larger
number of stars, including fainter stars.
Han (1997) estimated that the increase in $\tau$ caused by the miscount of 
effectively monitored stars is large enough to compensate the decrease 
in $\tau$ due to the decrease in timescales and make the observed optical
depth overestimated by a factor $\sim 1.7$.

   In addition to background stars the lens itself can cause blending
known as ``bright lens blending'' (Kamionkowski 1995; Buchalter \&
Kamionkowski 1997; Nemiroff 1997; Han 1998a).  Bright lens blending causes the
measured timescale and optical depth to be underestimated in the same manner
as regular blending.  In addition to the effects of
regular blending, lens blending causes the determined optical depth to
depend on the lens location.  This is because detecting events caused by
bright lenses close to the observer is comparatively more difficult than
detecting events produced by lenses near the source.  Additionally, since the
brighter the lens is, the more massive it tends to be, and thus the more
likely it is to be affected by lens blending.  As a result, the decrease in
optical depth for massive lenses is relatively bigger than the decrease for
low-mass lenses, making the measured optical depth dependent on the lens mass
function.

   Finally, the last type of blending, known as ``binary star blending'' 
occurs when the source is a binary.  In general, the typical separation 
between the component stars in a binary system is larger than the typical 
size of the Einstein ring radius.  In this case only one star is significantly
amplified, and the flux from the other star simply contributes as
blended light (Dominik 1998).  For some binaries with small component
separations, on the other hand, both component stars can be amplified.  
However, even in these cases it is not easy to detect the binarity of the 
source because the light curve mimics that of a single source event with 
a longer timescale and larger impact parameter than the true values 
(Griest \& Hu 1992; Han \& Jeong 1998).

\section{Limitations of Current Blending-Correction Methods}

   There have been various methods proposed to correct for the blending problem.
However, these methods are either applicable to only a few specific
types of blending or impractical due to the large uncertainty in the 
determined value of $f$.  In this section, we list these proposed methods 
and discuss their performance in correcting blending effects.

   Because a blended event light curve is not exactly the same as that of 
an unblended event, ideally the effects are detectable using {\it
photometry}.  However, due to the limits on photometric precision of the
current lensing experiments, blending has been photometrically detected only
in a small number of events.  Nevertheless, photometry is still being
discussed because of the rapidly increasing photometric precision made
possible through the operation of early warning systems (MACHO: Alcock et
al.\ 1996; OGLE: Udalski et al.\ 1994) and subsequent effective followup
observations (PLANET: Albrow et al.\ 1998; GMAN: Alcock et al.\ 1997c).  
However, we still find that even with this higher precision photometry, the 
uncertainties in the derived lensing parameters are very large, 
making it difficult to properly correct for blending
effects.  To demonstrate this, we simulate model events that are affected
by various amounts of blended light and estimate the uncertainties of the
recovered lensing parameters.  The events are assumed to have $\beta=0.3$,
and $t_{\rm E,0} = 15\ {\rm days}$, which are the most common values for the
Galactic bulge events, and are affected by blending with various blended
light fractions of $f=0.3$, 0.5, 0.7, and 0.9.  The events are assumed to be
observed 5 times/day during $-0.5 t_{\rm E}\leq t_{\rm obs}\leq 3t_{\rm E}$
with a high photometric precision of $p=1\%$.  The light curves are then
fit with a theoretical blended light curve as given by equation (1.3).  We
estimate the uncertainties of the recovered lensing parameters by 
computing the values of $\chi^2$ by
$$
\chi^2 = \sum_{i=1}^{N_{\rm dat}} 
\left( {A_{O}-A_{T} \over pA_{T}}\right)^2,
\eqno(3.1)
$$
where $N_{\rm dat}$ is the number of data points and $A_O$ and $A_T$
represent the simulated and theoretical model light curves, respectively.  
In Figure 1, we present the resulting values of $\chi^2$ 
as functions of $t_{\rm E}/t_{\rm E,0}$ and $f$.  The degrees of
freedom are determined by ${\rm dof}=N_{\rm dat}-N_{\rm par}-1$, where $N_{\rm
par}=4$ is the number of lensing parameters for the blended light curve fit.
The contours are drawn at the levels of $\Delta\chi^2=1.0$, 4.0, and 9.0
(i.e., $1\ \sigma$, $2\ \sigma$, and $3\ \sigma$ levels) from inside to
outside.  From the figure, one finds that the uncertainties in both the
derived values of $t_{\rm E}$ and $f$ are large.  One also finds that the
uncertainty ranges of the blended light fraction measured at all levels
include $f=0$ (i.e., no blended light), implying that it will be difficult to
detect blending effect even with high precision photometry.  In
addition, since the uncertainties increase rapidly with increasing values of
$f$, correcting for blending effects using this method will be especially
difficult for amplification-bias events which are severely affected by
blending.

   Another method of correcting for blending effects is to detect color shifts
during an event with high precision, multi-band photometry (Buchalter,
Kamionkowski, \& Rich 1996).  This method might be applicable if the lensed
source star has a very different color from that of the integrated light of
the blended stars.  However, the expected amount of color shift is generally
very small because most Galactic bulge stars have similar colors (Goldberg
1998).  In addition, even when the color shift is detected, the determined
lens parameters suffer from the same degeneracy as in the case of single-band
photometry, making it difficult to properly correct for the blending effect
(Wo\'zniak \& Paczy\'nski 1997).

   A more general method of correcting for blending effects is by 
detecting the shifts of a source star's image centroid, $\delta x$, during 
an event (Alard, Mao, \& Guibert 1995; Alard 1996).  If one of many stars 
in a blended seeing disk is gravitationally amplified, the position of 
the center of light will shift toward the lensed star by an amount
$$
\delta x = 
\eta \left\vert \langle \vec{x}\rangle -\vec{x}_0\right\vert;
\qquad \eta = {A_{\rm obs}-1\over A_{\rm obs}}
= {(1-f)(A_{\rm abs}-1)\over A_{\rm abs}(1-f) + f},
\eqno(3.2)
$$
where $\langle \vec{x}\rangle$ is the position of the center of light before
gravitational amplification and $\vec{x}_0$ is the location of the lensed
star.  Goldberg \& Wo\'zniak (1997) actually applied this method to the OGLE
data base and found that 7 out of 15 tested events showed significant
centroid shifts of $\delta x \gtrsim 0''\hskip-2pt .2$, demonstrating the
usefulness of this method.  However, not all blended events produce large
centroid shifts.  If the amplification of an event is very low, i.e., $A_{\rm
abs}\sim 1$, the expected centroid shift is small since $\eta\sim 0$.  In
addition, if the lensed star is the brightest one in the blended seeing disk
and its flux dominates that from the other blended stars, the expected amount
of centroid shift is very small even for a high amplification event because
the position of the center of light before gravitational amplification will
be very close to that of the lensed star, i.e., $\left\vert \langle
\vec{x}\rangle -\vec{x}_0\right\vert \sim 0$.  For large centroid shifts,
therefore, the lensed star should be one of the faint stars in the seeing
disk so that it has a negligible effect on the position of the center of
light (Han, Jeong, \& Kim 1998).  For amplification-biased events, source
stars are generally very faint.  Because they are faint, the fact that they
are detectable implies that the source stars are highly amplified.  Since the
conditions for detecting amplification-biased events agree well with those
for large centroid shifts, the centroid shift measurement is an efficient
method for detecting amplification-bias blending.  However, this method is
not efficient for detecting other types of blending.  First, because of a
relatively small amount of blended light, regular blended events do not need
to be highly amplified to be detected.  Although they can be highly
amplified, the dominance of the source flux over that from other faint
blended stars will result in small centroid shifts.  
This method is also inefficient for an event affected by the light from very 
closely located blends such as binary companions and bright
lenses due to the small amount of expected centroid shift.

\section{Correction of Blending Effect by SIM}

   In the previous section, we showed that correcting for blending effects for
general microlensing events is difficult by using the previously proposed 
methods.  In this section we show that with diffraction-limited, single-mirror
images and the high astrometric precision of SIM, one can correct 
for the effects of nearly all types of blending.

  The optics of SIM can be modeled by a diffraction-limited telescope with 
two circular mirrors.  Then the expected PSF is given by the classical double 
circular-aperture diffraction pattern formula of
$$
I(\theta) = I_0
   \left[ {2J_1(\pi\theta D/\lambda) \over \pi\theta D/\lambda}\right]^2
            \cos^2 \left( {\pi\theta d\over \lambda}\right),
\eqno(4.1)
$$
where $I_0$ is the intensity of the central peak, $d$ is the separation 
between the two mirrors with an aperture diameter $D$, $\lambda$ is the 
observed wavelength, $\theta$ is the angle measured from the central fringe, 
and $J_1$ is the 1st-order Bessel function (Hecht 1998).  According to the 
specification of SIM, $d=10$ m and $D=0.3$ m.  The upper left two panels of 
Figure 2 show the expected PSF when a single (unblended) source star is 
observed by SIM.  When observed at $\lambda\sim 5500\ \AA$, the PSF has a 
``broad envelope'' of a width $1.22\lambda/D \sim 0.\hskip-2pt ''5$ modulated 
by ``narrow fringes'' of a width $\lambda/d \sim 10\ {\rm mas}$.

   When the source star is blended, however, the PSF takes a different form 
from that in equation (4.1).  The blended source star PSF can be classified 
into three regimes depending on the separation $\Delta\theta$ between the 
lensed source and the blended star: regime 1 ($\Delta\theta \gtrsim 
1.22\lambda/D$), regime 2 ($\Delta\theta < \lambda/d$), and regime 3 
($\lambda/d \leq \Delta\theta < 1.22\lambda/D$).  In the following subsections, we investigate how the blends with various separations affect the 
PSF and their effects can be corrected for individual cases.

\subsection{Regime 1 ($\Delta\theta \gtrsim 
	               1.22\lambda/D\sim 0.\hskip-2pt ''5$)}

   In this regime the broad envelopes of the two sources do not overlap
(see the lower left panels of Figure 2), and thus the blended source is easily
distinguished.  In Baade's Window the density of stars with $V < 23$, which are 
the major sources of blending, is less than $1\ {\rm arcsec}^{-2}$ according 
to the luminosity function determined from the {\it Hubble Space Telescope} 
observations by Holtzman et al.\ (1998).  The stellar density in LMC field 
is even smaller.  This implies that a significant fraction of sources blended 
by field stars, i.e., regular and amplification-bias blended events, are 
automatically resolved by a single mirror of SIM.

\subsection{Regime 2 ($\Delta\theta < \lambda/d\sim 10\ {\rm mas}$)}

   In this case the narrow fringes from the two images overlap (the upper
right panels of Figure 2).  The position of the centroid of the central 
fringe, which is the position to be astrometrically measured by SIM,
reflects the flux-weighted average of the two sources separately.
Two classes of blending sources belong to this regime: a large fraction 
of binary companions to the source and all bright lenses.  Of these two 
types of blending, Jeong, Han, \& Park (1999) discussed in detail 
the effects of bright lenses on SIM observations and how to correct 
for blending effects caused by bright lenses.  Here we focus on the 
effects of blending on events affected by binary star blending.

   When a source star is gravitationally amplified, it is split into 
two images located on the same and opposite sides of the lens, respectively
(see Figure 2 of Paczy\'nski 1996).  Due to the changes in position and
amplification of the individual images caused by the lens-source transverse
motion (see Figure 3 of Paczy\'nski 1996), the light centroid between the
images changes its location during the event.  The location of the image
centroid relative to the source star is related to the lensing parameters by
$$
\vec{\delta\theta_{c}} = {\theta_{\rm E}\over u^2+2}
({\cal T}\hat{\bf x} + \beta\hat{\bf y}),
\eqno(4.2.1)
$$
where ${\cal T}=(t-t_0)/t_{\rm E}$ and $\hat{\bf x}$ and $\hat{\bf y}$ are
the unit vectors toward the directions which are parallel and normal to the 
lens-source transverse motion, respectively.  If we let 
$(x,y)=(\delta\theta_{c,x}, \delta\theta_{c,y}-b)$ and 
$b=\beta\theta_{\rm E}/2(\beta^2+2)^{1/2}$, 
the coordinates are related by
$$
	x^2 + {y^2\over q^2} = a^2,
\eqno(4.2.2)
$$
where $a = \theta_{\rm E}/2(\beta^2+2)^{1/2}$ and 
$q=b/a=\beta/(\beta^2+2)^{1/2}$.  Therefore, during the event the trajectory
of the source star image centroid traces out an astrometric ellipse 
(Walker 1995; Boden et al.\ 1998; Jeong et al.\ 1999).

   However, when an event is affected by binary star blending, the centroid 
shift is affected by the light from the companion star and its trajectory 
deviates from an ellipse.  Therefore, by detecting the distortion of the 
astrometric shift trajectory, it is possible to identify the blend and thus 
correct for the blending effect.  In Figure 3, we illustrate the distortion 
of the astrometric shift trajectory when an event is blended by a binary 
companion with a light fraction of $f=0.3$.  In the figure, the dotted line 
represents the unperturbed trajectory of the centroid shift with respect to 
the position of the lensed source star located at the origin.  As mentioned, 
the trajectory is an ellipse.  On the other hand, when the event is blended 
by a companion star, located at $\vec{\delta\theta}_B$ with respect to the 
lensed star, the position of the light centroid will shift toward the 
companion.  In addition, the reference position of the astrometric 
measurements is not the position of the lensed source star but the 
center of light between the binary components before amplification.  Due to 
these two effects, the resulting trajectory (represented by a solid line) of 
the astrometric shifts is no longer an ellipse.  Note that since the majority 
of binaries in this regime have long orbital periods, for example 
$P\sim 30\ {\rm yrs}$ for a Galactic bulge binary system with 
$\Delta\theta\sim 1\ {\rm mas}$, the motion of the blend is not important.  
Assuming that the distribution of separations for binaries in the solar 
neighborhood determined by Duquennoy \& Mayor (1991) can be applied 
to bulge and LMC populations, $\gtrsim 60\%$ of Galactic and $\gtrsim 70\%$ 
of LMC binaries belong to this regime.

   The trajectory of the centroid shift for a blended, binary event takes 
various forms depending on the fraction of blended light and the location of 
the blended star with respect to the lensed star.  The centroid shift for 
an event blended by a binary companion is represented by
$$
\vec{\delta\theta}_{c,{\rm obs}} =
{1-f \over f+A_{\rm abs}(1-f) }
\left[ A_{\rm abs}\vec{\delta\theta}_c -
f\vec{\delta\theta}_B (A_{\rm abs}-1)\right].
\eqno(4.2.3)
$$
In the equation, the term including $\vec{\delta\theta}_c$ describes the
elliptical motion of the centroid shift with respect to the lensed star (the
elliptical term).  On the other hand, the term including
$\vec{\delta\theta}_B$ describes the linear shift caused by the light from
the blended star (the linear term).  Therefore, the shape of the astrometric
shift trajectory for a blended event results from the combination of the
elliptical displacement caused by gravitational lensing and the linear
displacement toward the blended star.
In Figure 4, we present various forms of the centroid shift trajectory
as distorted by binary star blending.  The left panels show how the
trajectory changes from the unperturbed astrometric ellipse (in the top
panel) with increasing fractions of blended light.
To see the variation in the trajectory with respect to the location of the 
companion star, we also present the trajectories for various 
binary separations in the right panels.

\subsection{(Regime 3 $\lambda/d \leq \Delta\theta < 1.22\lambda/D$)}

   In this case both the broad envelopes and fringes of the two PSFs
are not matched (the lower right panels of Figure 1), and thus both 
centroids are shifted.  The position of the centroid between the 
central broad envelopes is the flux-weighted average of the two sources 
and the amount of the shift is given by equation (3.2).  One way to detect 
and correct for blending in this regime, therefore, is to measure the shift of 
the broad envelope centroid.  However, the situation is slightly more complex 
for the position of the centroid of the narrow fringes.  If the separation 
between two source is exactly the same as $n$ times the width of the 
narrow fringe, i.e., $\Delta\theta = n\lambda/d$, the 0th fringe of the 
blend coincide exactly with the $n$th fringe of the lensed source.  
However, this is true only at a specific wavelength.  If the blended
source is observed at twice as long a wavelength, for example, the 0th fringe
of the blend coincides exactly with the 1st null of the lensed source.
Therefore, from the difference in the interference patterns observed at
multiple wavelengths, one can deduce the position and brightness of the
blend even if the observer chooses to examine the central fringe rather than
the whole interference pattern.  The sources of blending belonging to this 
regime include large-separation binary companions to the lensed source and 
very closely located field stars.

\section{Summary}

   By observing gravitational microlensing events with SIM, one can correct 
for the effects of nearly all types of blending.  With its diffraction-limited, 
$\lesssim 0.\hskip-2pt ''5$ imaging, a single mirror of SIM can resolve blended 
stars for a significant fraction of the events blended by field stars, such 
as regular and amplification-bias blended events.  
For events affected by very close blends with 
separations $\Delta\theta\lesssim 10\ {\rm mas}$, such as a large fraction of 
binary-star blended events, it is also possible to identify the blend by 
detecting a distortion in the astrometric shift trajectory.  For events 
affected by blends with medium range separations, such as closely located 
field stars and wide binary companions, one can correct for the effects of 
blending by either detecting the shift of the broad envelope centroid between 
the PSFs of lensed and blended stars or by examining the narrow fringe 
patterns observed at multiple wavelengths.

\acknowledgements
We would like to thank to M.\ Everett and P.\ Martini for a careful reading 
of the manuscript.  
We also would like to thank the referee (A.\ Gould) for useful suggestions
that improved the paper.


\clearpage

\begin{figure}[t]
\centerline{\hbox{\epsfysize=16truecm \epsfbox{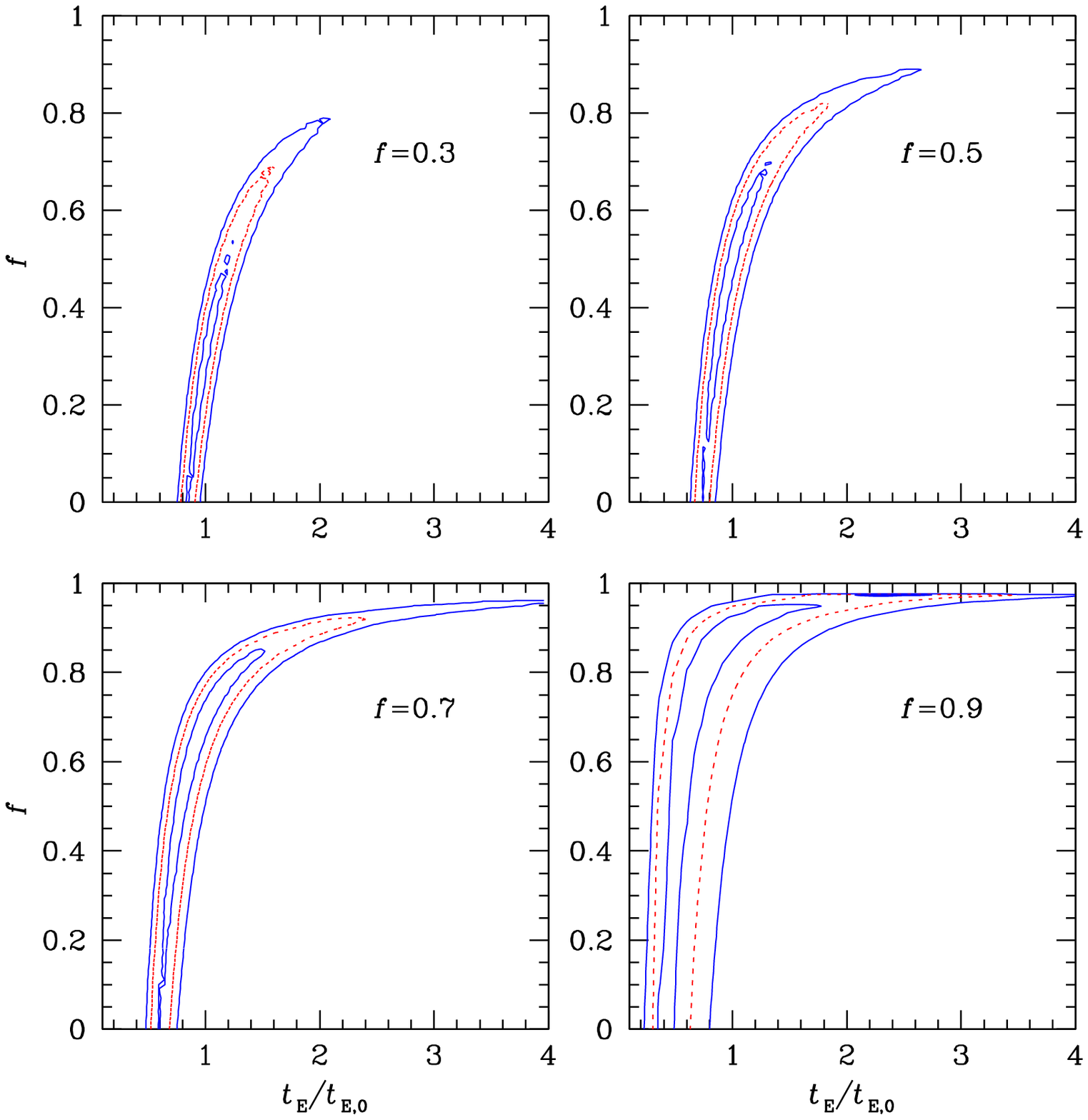}}}
\noindent
{\small {\bf Figure 1:}\
The uncertainties in the lensing parameters for events whose high precision 
light curves are fit with models.  The events
are assumed to have $t_{\rm E}=15\ {\rm days}$ and $\beta=0.3$, and are
affected by blending with blended light factions of $f=0.3$, 0.5,
0.7, and 0.9.  The events are assumed to be observed 5 times/day during
$-0.5t_{\rm E}\leq t_{\rm obs}\leq 3t_{\rm E}$ with a photometric
precision of $p=1\%$.  The uncertainties in the lensing parameters are
determined by computing $\chi^2$, and the resulting $\chi^2$ as
functions of $t_{\rm E}/t_{\rm E,0}$ and $f$ are presented as contour maps.
In each panel, the contours are drawn at $1\sigma$, $2\sigma$, and $3\sigma$
levels from inside to outside.
}
\end{figure}

\begin{figure}[t]
\centerline{\hbox{\epsfysize=16truecm \epsfbox{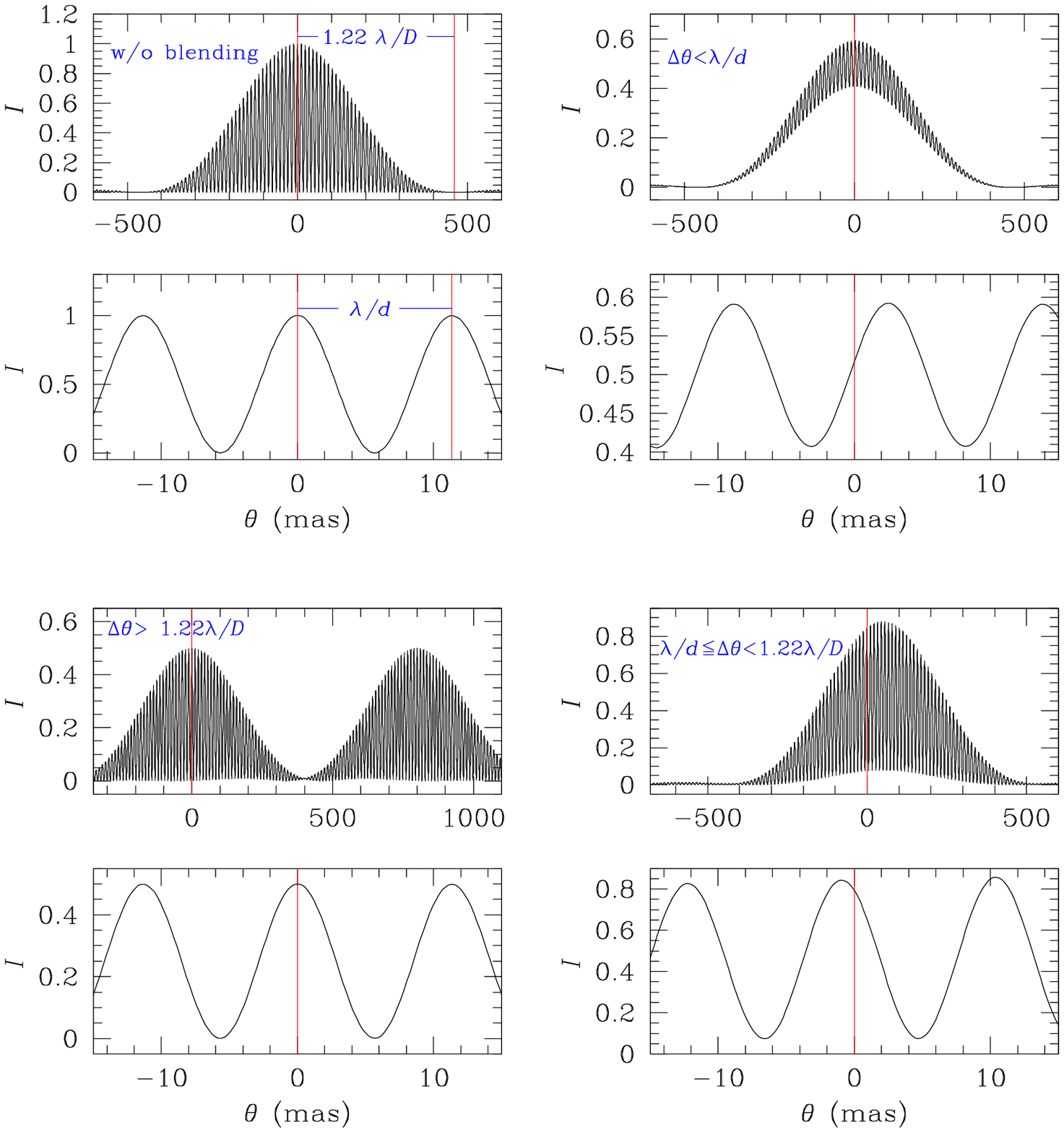}}}
\noindent
{\small {\bf Figure 2:}\
The expected fringe patterns when a source star affected by a blended star 
with various separations is observed by SIM.  To better show the centroid shift 
of the narrow fringe pattern, the region around the position of the central 
fringe of the lensed star is expanded in the lower panel of each PSF.  
For all of these model events, we assume a blended light fraction of 
$f=0.5$ and arbitrarily normalize the intensity. 
}
\end{figure}

\begin{figure}[t]
\centerline{\hbox{\epsfysize=16truecm \epsfbox{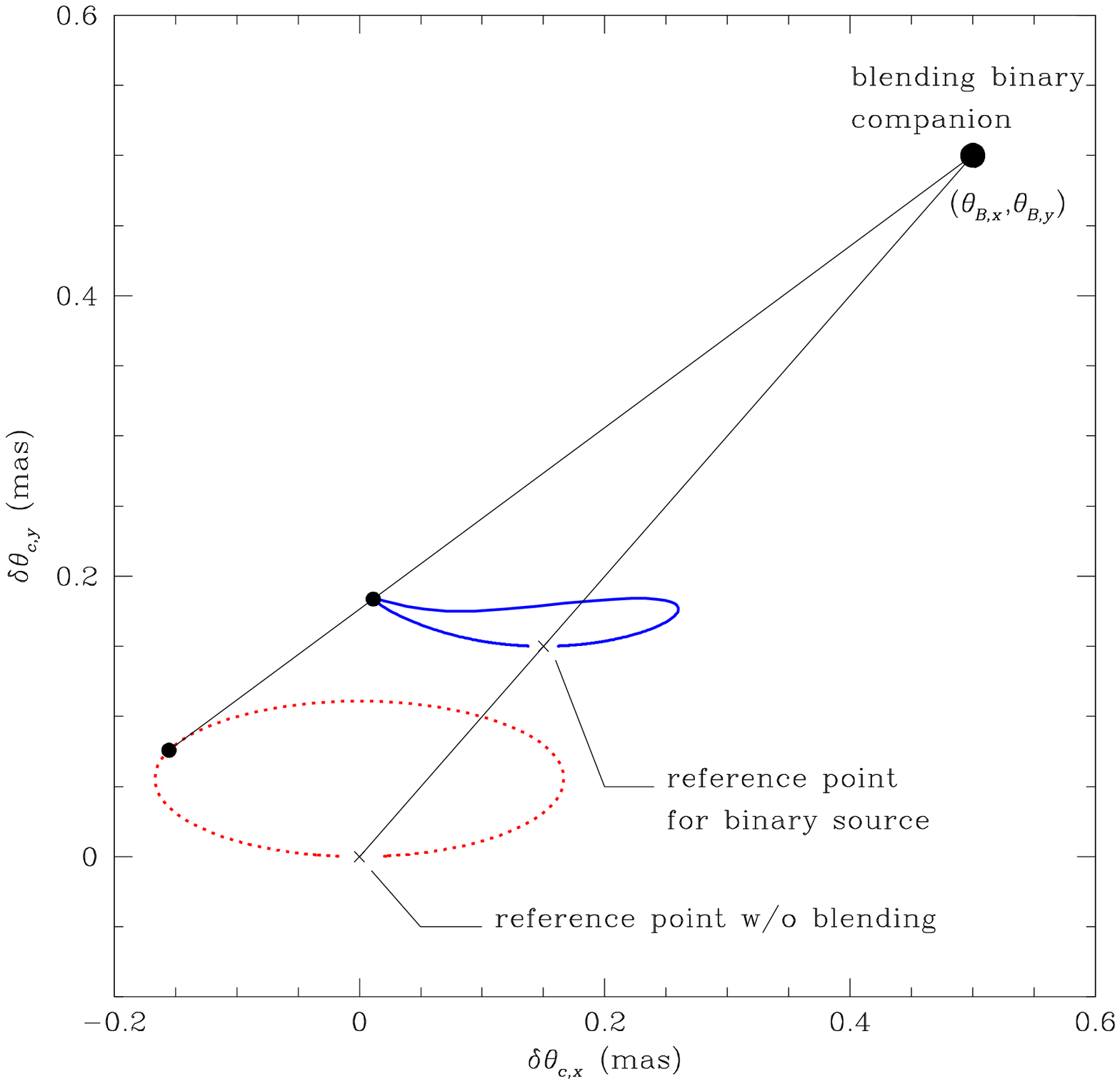}}}
\noindent
{\small {\bf Figure 3:}\
Illustration of the distortion of the astrometric shift trajectory by a binary
companion.  In the figure, the dotted line represents the trajectory of
the centroid shifts with respect to the position of the source star located
at the origin when the star is not affected by the binary-star blending.  
On the other hand, when the event is blended by a companion star, located at
$(\theta_{B,x},\theta_{B,y})$, the position of the light centroid will shift
toward the companion.  In addition, the reference position of the astrometric 
measurements is not the position of the lensed source star, but the center 
of light between the component stars.  Due to these combined effects, the 
resulting trajectory (represented by a solid line) of the astrometric shifts 
is no longer an ellipse.  This model event has lensing parameters of 
$t_{\rm E}=15\ {\rm days}$, $\theta_{\rm E}=0.5\ {\rm mas}$, and $\beta=0.5$.  
The light fraction of the blended star is $f=0.3$.
}
\end{figure}

\begin{figure}[t]
\centerline{\hbox{\epsfysize=16truecm \epsfbox{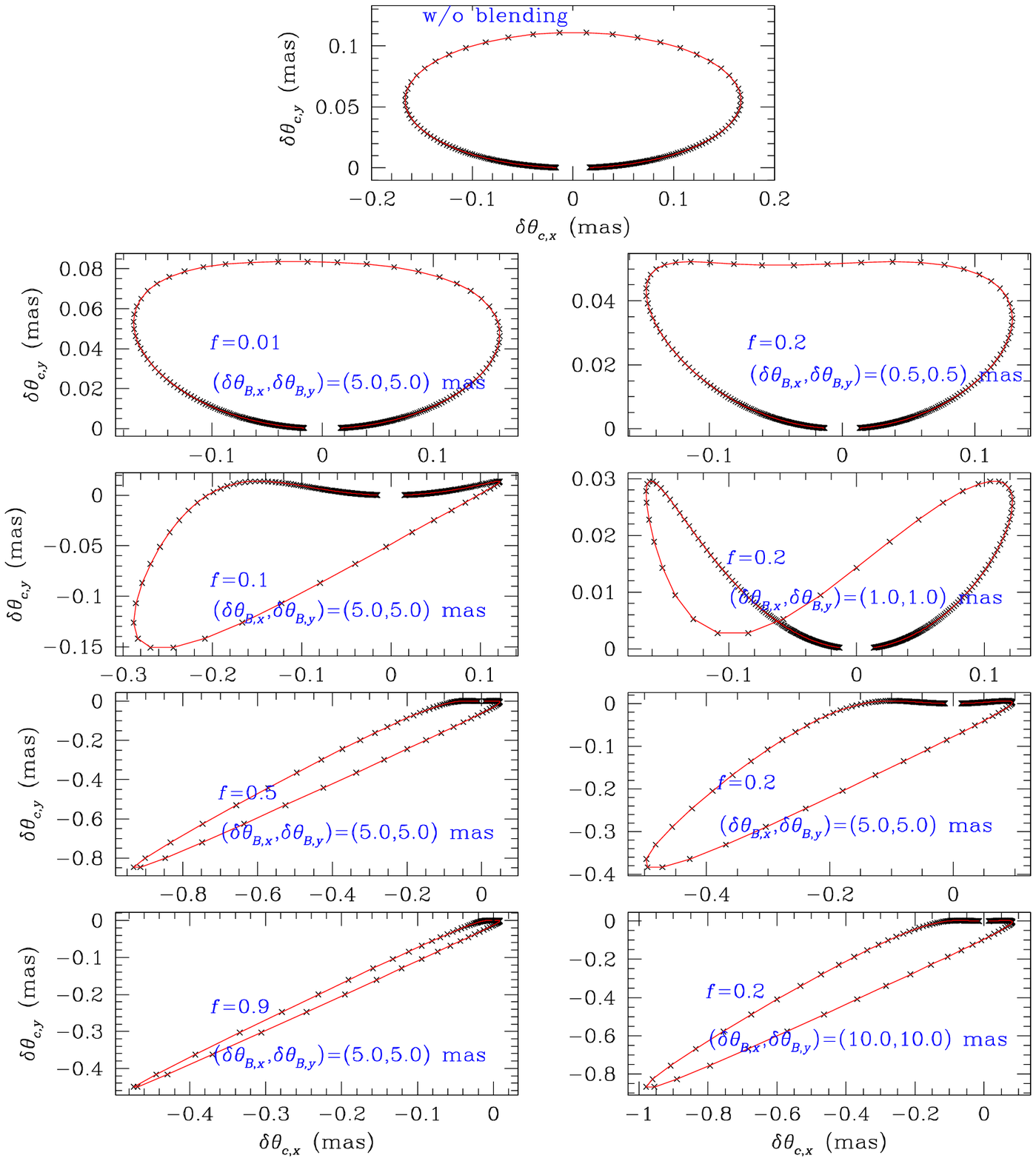}}}
\noindent
{\small {\bf Figure 4:}\
Various forms of the centroid shift trajectory distorted by binary-star
blending.  The left-side panels show how the trajectory changes from the 
unperturbed astrometric ellipse (in the top panel) with increasing fractions 
of blended light.  To see the variation of the trajectory with respect to 
the location of the companion star, we also show the trajectories for 
various binary separations in the right-hand panels.  This model event 
has lensing parameters of $t_{\rm E}=15\ {\rm days}$, 
$\theta_{\rm E}=0.5\ {\rm mas}$, and $\beta=0.5$.
}
\end{figure}


\clearpage
\begin{references}
\reference{alard96}         Alard, C.\ 1996, in IAU Symp.\ 173,
                            Astrophysical Applications of Gravitational
                            Lensing, ed.\ C.\ S.\ Kochanek \& J.\ N.\
			    Hewitt (Dordrecht: Kluwer), 215
\reference{alard97}         Alard, C. 1997, \aap, 321, 424
\reference{alard95}         Alard, C., Mao, S., \& Guibert, J.\ 1995,
	                    \aap, 300, L17
\reference{albrow98}        Albrow, M., et al.\ 1998, \apj, 509, 000 
\reference{alcock96}        Alcock, C., et al.\ 1996, \apj, 463, L67
\reference{alcock97a}       Alcock, C., et al.\ 1997a, \apj, 479, 119
\reference{alcock97b}       Alcock, C., et al.\ 1997b, \apj, 486, 697
\reference{alcock97c}       Alcock, C., et al.\ 1997c, \apj, 491, 436 
\reference{allen98}         Allen, R., Shao, M., \& Peterson, D.\ 1998,
		            Proc.\ SPIE, 2871, 504
\reference{ansari96}        Ansari, R., et al.\ 1996, \aap, 314, 94
\reference{boden98}         Boden, A.\ F., Shao, M., \& 
			    Van Buren, D.\ 1998, \apj, 502, 538          
\reference{bouquet93}       Bouquet, A.\ 1993, \aap, 280, 1
\reference{buchalter96}     Buchalter, A., Kamionkowski, M., \& Rich, M.\ R.\
                            1996, \apj, 469, 676
\reference{buchalter97}     Buchalter, A., \& Kamionkowski, M.\ 
			    1997, \apj, 482, 782
\reference{distefano95}     Di Stefano, R., \& Esin, A.\ A.\ 1995, 
			    \apj, 448, L1    
\reference{dominik98}       Dominik, M.\ 1998, \aap, 333, 893
\reference{duquennoy91}     Duquennoy, A., \& Mayor, M.\ 1991, \aap, 248, 485
\reference{goldberg98}      Goldberg, D.\ M.\ 1998, \apj, 489, 156
\reference{goldberg97}      Goldberg, D.\ M., \& Wo\'zniak, P.\ R.\ 1998, 
			    Acta Astron., 48, 19
\reference{griest92}        Griest, K., \& Hu, W.\ 1992, \apj, 397, 362
\reference{han97}           Han, C.\  1997, \apj, 484, 555
\reference{han98a}          Han, C.\ 1998a, \apj, 500, 569
\reference{han98b}          Han, C.\ 1998b, \mnras, submitted
\reference{han98c}          Han, C., \& Chang, K.\ 1998, \mnras, submitted
\reference{han98c}          Han, C., \& Jeong, Y.\ 1998, \mnras, 301, 231
\reference{han98d}          Han, C., Jeong, Y., \& Kim, H.-I.\ 1998, 
			    \apj, 507, 102
\reference{hecht98}         Hecht, E.\ 1998, Optics 
			    (Rending: Addison-Wesley Longman), 461
\reference{holtzman98}      Holtzman, J.\ A., et al.\ 1998, \aj, 115, 1946
\reference{jeong99}         Jeong, Y., Han, C., \& Park, S.-H.\ 1999,
			    \apj, 511, 000
\reference{kamionkowski95}  Kamionkowski, M.\ 1995, \apj, 442, L9
\reference{nemiroff94}      Nemiroff, R.\ J.\ 1994, \apj, 435, 682
\reference{nemiroff97}      Nemiroff, R.\ J.\ 1997, \apj, 486, 693
\reference{Paczynski96}     Paczy\'nski, B.\ 1996, \araa, 34, 419
\reference{Paczynski98}     Paczy\'nski, B.\ 1998, \apj, 404, L23
\reference{udalski94}       Udalski, A., et al.\ 1994, Acta Astron.,  
                            44, 227
\reference{udalski97}	    Udalski, A., et al.\ 1997, Acta Astron., 
		            47, 169
\reference{walker95}        Walker, M.\ A.\ 1995, \apj, 453, 37
\reference{wozniak97}       Wo\'zniak, P., \& Paczy\'nski, B.\ 1997, 
			      \apj, 487, 55.
\end{references}
\end{document}